# Energy-resolved Photoconductivity Mapping in a Monolayer-bilayer WSe$_2$ Lateral Heterostructure

Zhaodong Chu[1†], Ali Han[2†], Chao Lei[1], Sergei Lopatin[3], Peng Li[2], David Wannlund[1], Di Wu[1], Kevin Herrera[1], Xixiang Zhang[2], Allan H. MacDonald[1], Xiaoqin Li[1], Lain-Jong Li[2], Keji Lai[1]*

[1]*Department of Physics, Center for Complex Quantum Systems, The University of Texas at Austin, Austin, Texas 78712, USA*

[2]*Physical Sciences and Engineering Divison, King Abdullah University of Science and Technology, Thuwal 23955-6900, Kingdom of Saudi Arabia*

[3]*King Abdullah University of Science and Technology (KAUST), Core Labs, Thuwal, 23955-6900, Saudi Arabia*

[†] These authors contributed equally to this work

* E-mails: kejilai@physics.utexas.edu

# Abstract

Vertical and lateral heterostructures of van der Waals materials provide tremendous flexibility for band structure engineering. Since electronic bands are sensitively affected by defects, strain, and interlayer coupling, the edge and heterojunction of these two-dimensional (2D) systems may exhibit novel physical properties, which can be fully revealed only by spatially resolved probes. Here, we report the spatial mapping of photoconductivity in a monolayer-bilayer $WSe_2$ lateral heterostructure under multiple excitation lasers. As the photon energy increases, the light-induced conductivity detected by microwave impedance microscopy first appears along the hetero-interface and bilayer edge, then along the monolayer edge, inside the bilayer area, and finally in the interior of the monolayer region. The sequential emergence of mobile carriers in different sections of the sample is consistent with the theoretical calculation of local energy gaps. Quantitative analysis of the microscopy and transport data also reveals the linear dependence of photoconductivity on the laser intensity and the influence of interlayer coupling on carrier recombination. Combining theoretical modeling, atomic scale imaging, mesoscale impedance microscopy, and device-level characterization, our work suggests an exciting perspective to control the intrinsic band-gap variation in 2D heterostructures down to the few-nanometer regime.

KEYWORDS: Van der Waals materials, monolayer-bilayer interface, edge states, microwave impedance microscopy, photoconductivity imaging

The electronic properties of two-dimensional (2D) van der Waals (vdW) materials are highly sensitive to the local atomic arrangement[1,2], which offers tremendous opportunities for band structure engineering. In transition metal dichalcogenides (TMDs), for instance, the band gap can be tuned by choosing different $MX_2$ (M = Mo, W…; X = S, Se, Te) compounds and alloys[3,4], changing the number of layers[5,6], vertical stacking or lateral stitching of different materials[1,2,7], and in the case of vertical heterostructures, varying the twist angle between the atomic layers[8,9]. Moreover, broken crystalline symmetry at the edges and hetero-interfaces of vdW materials may lead to the emergence of new electronic states in the band structure[10,11], which could even become topologically nontrivial in certain cases[12-14]. Taken into account these possibilities, the local band structure in 2D materials and heterostructures can be readily explored for electronic applications, provided such spatial variations are well understood and controlled[15-17].

Many optoelectronic applications such as photodetectors and photovoltaic devices require photo-excitation of mobile carriers by above-gap illumination and the subsequent charge transport[18]. In conventional photocurrent spectroscopy (PCS) experiments[19], the sample is illuminated by a wavelength-tunable light source and the induced photocurrent is measured as a function of the photon energy. For spatially resolved studies, scanning photocurrent microscopy (SPCM) with a diffraction-limited resolution of 0.5 ~ 1 μm can map out the photo-generated current across the source and drain electrodes when a focused laser beam scans over the device[20]. The SPCM signal is usually strongest near the electrodes because the current is most effectively generated in regions with a large electric field. Its interpretation can be further complicated by carrier diffusion, Schottky contact, thermoelectric and bolometric effects[20-23], making it unsuitable to investigate spatial variations of intrinsic material properties. For thin films deposited on metallic substrates, the light-induced conductance can also be detected by conductive atomic-force microscopy (C-AFM)[24], which again suffers from the extrinsic Schottky effect. Here, we report the imaging of intrinsic photoconductivity in a monolayer-bilayer (ML-BL) $WSe_2$ lateral heterostructure by microwave impedance microscopy (MIM)[25] with light stimulation at multiple wavelengths. Taking advantage of the capacitive coupling, the MIM can perform noninvasive nanoscale photoconductivity imaging without the need for electrical contacts[26,27]. As the photon energy of excitation lasers increases from 1.16 eV to 2.78 eV, the photo response first appears along the ML-BL interface and the BL edge, then along the ML edge and inside the BL bulk, and finally in the interior of the ML. The spatial evolution of photo response is consistent with the variation of local band gaps calculated by the first-principles method. The mesoscale photoconductivity mapping fills an important void between atomic scale characterization and device-level transport measurements. Our results suggest the exciting possibility of controlled in-plane band-gap engineering on the length scale of nanometer, which is uniquely enabled by atomically thin lateral heterostructures and may lead to completely new integrated devices.

The $WSe_2$ nano-flakes were grown on double-side polished sapphire substrates by chemical-vapor deposition (CVD). Fig.1a shows an optical microscopy image of the as-grown flakes, where the monolayer and bilayer regions display different optical contrast. In this work, we focus on flakes with both regions, i.e., on a ML-BL lateral heterostructure whose layer thickness was confirmed by the AFM image in Fig. 1b. The Raman and photoluminescence (PL) spectra of the ML and BL regions (Figs. 1c and 1d) were measured by a confocal Raman microscope. The Raman intensities of both the in-plane $E^1_{2g}$ (~128 cm$^{-1}$) and out-of-plane $A_{1g}$ (~246 cm$^{-1}$) phonon modes in bilayer $WSe_2$ are weaker than that in the monolayer, whereas the out-of-plane $B^1_{2g}$ (~304 cm$^{-1}$) mode only appears in the bilayer region, consistent with previous investigations[6,28]. The Raman maps by integrating these three modes are shown in Supporting Information Figure S1. The PL signal of monolayer $WSe_2$ at ~780 nm is much stronger than that of the bilayer area at ~794 nm due to the direct-to-indirect-gap transition[5,6,29,30]. In order to determine the atomic arrangement of the sample, we carried out high-resolution imaging by transmission electron microscopy in the scanning mode with high-angle-annular dark-field (HAADF-STEM)[31] around the monolayer edge (Fig. 1e), hetero-interface (Fig. 1f) and bilayer edge (Fig. 1g). The ADF-STEM images show the expected hexagonal structures in the interior of the flakes, with clear contrast between the ML and BL regions. The ML-BL heterojunction and sample edges display a mixture of zigzag and armchair patterns, as well as many defective bonds. At the atomic length scale, therefore, these CVD-grown $WSe_2$ lateral heterostructures are characterized by ordered 2D crystals in the monolayer/bilayer regions and disordered 1D edges and interfaces.

We now investigate the intrinsic band-gap variation in the as-grown ML-BL $WSe_2$ sample on sapphire substrates using light-stimulated MIM[26,27], which is based on a commercial AFM platform (Park Systems XE-100). In this work, several diode lasers (pigtail laser diodes from Thorlabs, Inc.) with wavelengths ranging from 1065 nm to 446 nm are guided by a single-mode optical fiber to illuminate the flakes through the transparent sapphire substrate, as illustrated in Fig. 2a. During the experiments, the center of the focused laser spot (~ 10 μm in diameter for infrared and 4 ~ 6 μm for the visible light) is aligned with the MIM tip and the sample stage is scanned in the *xy*-plane. The 1 GHz microwave signal is fed to the center conductor of the shielded cantilever probe[32]. The photo-induced charge carriers can screen the GHz electric fields, which modifies the effective sample impedance. The MIM electronics then detect the real and imaginary parts of the tip-sample admittance to form the MIM-Re and MIM-Im images, respectively[33]. Note that the laser may simultaneously generate free carriers, excitons, trions, and other quasi-particles. However, since excitons are charge neutral and trions are insignificant in intrinsic semiconductors, their contribution to the MIM signals is negligible. In addition, the high-frequency coupling between the tip and sample does not require direct contact or source-drain electrodes, eliminating the problem of Schottky barriers in C-AFM and SPCM. To avoid band bending due to the metal-semiconductor contact, we also deposited a 30-nm $Al_2O_3$ layer on the $WSe_2$ flakes by atomic-layer deposition (ALD)[26,27,34].

A sequence of MIM images of the ML-BL $WSe_2$ flake in Fig. 1b, both in the dark and under illumination with different photon energies (power density $P \sim 2 \times 10^5$ mW/cm$^2$), are shown in Fig. 2b. As detailed in the quantitative analysis below, the signals in both MIM channels increase monotonically as a function of the photoconductivity around this laser intensity. The dark spots and lines in the MIM-Im images are due to the topographic crosstalk. For the longest wavelength ($\lambda = 1065$ nm, photon energy $E = 1.16$ eV) laser, there is virtually no change in the MIM images compared to that in the dark. In other words, the photoconductivity under this photon energy is below the sensitivity (see analysis below) of our technique. At a slightly shorter $\lambda$ of 979 nm ($E = 1.27$ eV), the ML-BL interface and the BL edge start to exhibit weak MIM-Re signals, which continue to increase in strength as $\lambda$ decreases to 915 nm (1.36 eV) and 829 nm (1.50 eV). The photo-excited carriers are spatially confined in the lateral interface or bilayer edges. The full-width-half-maximum (FWHM) of these quasi-1D conductive lines is ~ 200 nm (Supporting Information Fig. S2), which is limited by the tip diameter rather than the actual width of the edge states[25]. As the wavelength of the light source enters the visible regime ($\lambda = 699$ and 638 nm, $E = 1.77$ and 1.94 eV), MIM-Re signals also appear at the ML edge and inside the BL area. Finally, for the two shortest wavelength ($\lambda = 517$ and 446 nm, $E = 2.40$ and 2.78 eV) excitation lasers, photoconductivity is observed everywhere in the sample, with much higher signals in the BL than the ML region. The complete MIM images with various power densities are included in Supporting Information Fig. S2. The presence of a cutoff energy for each edge or bulk region is seen in the data. For instance, no MIM signal is detected at the ML edge under the 1.50 eV excitation even at a high power of $P \sim 10^6$ mW/cm$^2$, whereas the signal appears under the 1.77 eV excitation at a low $P \sim 10^4$ mW/cm$^2$ and increases as a function of the laser power. Similar results on a different flake can be found in Supporting Information Fig. S3.

The sequentially emerging MIM-Re signals across five different regions of the sample (along the white dashed line in Fig. 2b) are summarized in Fig. 2c. Photoconductivity at the bilayer edge and ML-BL heterojunction emerges for a photon energy between 1.16 and 1.27 eV. The effective local energy gap of these disordered quasi-1D channels is therefore ~ 1.2 eV. The gap changes to between 1.50 and 1.77 eV at the monolayer edge and bilayer bulk, and further increases to between 1.94 and 2.40 eV in the monolayer WSe$_2$. Micrometer-sized fluctuations of MIM signals can be seen inside the ML and BL regions, which are likely due to the spatial variation of carrier mobility due to defects and strain field. The wavelength-dependent photoconductivity imaging allows us to construct a qualitative band configuration in Fig. 2d. Because of the discrete excitation lasers in this experiment, the gaps are only determined within a range of energies. Future work using a light source with continuously tunable wavelengths will allow a more precise measurement on the nanoscale energy gaps. In a recent scanning tunneling microscopy/spectroscopy (STM/S) study on a similar CVD-grown WSe$_2$ sample on graphite[35], the band gaps at the ML-BL interface, BL bulk, and ML bulk were measured to be 0.8 eV, 1.8 eV, and 2.5 eV, respectively, which are in reasonable agreement with our findings.

Quantitative understanding of the MIM data can be obtained by finite-element analysis (FEA) of the tip-sample interaction[33]. We first discuss the 1D features in the sample, which are modeled as narrow nanoribbons in the simulation. Details of the FEA modeling are presented in Supporting Information Fig. S4. Note that the simulation results are invariant with respect to the 1D conductance $\sigma_{1D} = \sigma_{3D} \cdot t \cdot w$, where $\sigma_{3D}$, $t$, and $w$ are the 3D conductivity, thickness, and width of the conductive channel, respectively. As shown in Fig. 3a, the MIM-Im signal (proportional to the tip-sample capacitance) increases with increasing $\sigma_{1D}$ and saturates at both insulating and conducting limits, whereas the MIM-Re signal (proportional to the electrical loss) peaks at an intermediate $\sigma_{1D} \sim 10^{-13}$ S·m. Such response[33,34] is a consequence of the screening of microwave electric fields by mobile carriers in the sample, as shown in Fig. 3b. In Fig. 3c, we plot the power dependence of the MIM response of the ML edge illuminated by the 699 nm (1.77 eV) laser and the ML-BL interface and BL edge illuminated by the 915 nm (1.36 eV) laser. The wavelengths are chosen such that the nearby bulk regions remain insulating. The power-dependent photoconductivity can then be evaluated by comparing the simulation and experimental results. The linear $\sigma - P$ relation in Fig. 3d suggests that mobile carriers in the photo-induced conductive edges and interfaces follow the usual Drude behavior[18], i.e.,

$$\sigma = ne\mu_n + pe\mu_p = \eta \left[ (P\tau_n/h\nu)e\mu_n + (P\tau_p/h\nu)e\mu_p \right] \propto P \qquad (1)$$

where $\eta$ is the incident photo-to-electron conversion efficiency (IPCE), $n$ and $p$ are the photo-generated electron/hole density, $e$ is the electron charge, $\tau_n$ and $\tau_p$ are the electron/hole lifetime, $\mu_n$ and $\mu_p$ are the electron/hole mobility, and $h\nu = E$ is the photon energy. The roughly factor-of-two difference between photoconductivity in the ML and BL edges may be attributed to the difference in IPCE[36,37]. In other words, the recombination and scattering events are rather similar in these quasi-1D conductive channels.

To further investigate the hetero-interface and edge states in WSe$_2$, we have performed first-principles density functional theory (DFT) calculations. It is known that perfect zigzag edges in TMD materials are metallic in nature[10,38]. However, the actual edges and interfaces in our samples, as previously shown in the STEM images, consist of a disordered mix of zigzag, armchair, and bearded configurations. In this work, we calculate the semiconducting armchair edges in Fig. 4a, where the additional electronic states are evident inside the bulk gap. These topologically trivial states localized at the edges and interfaces are highlighted in the diagram in Fig. 4b. The calculated direct gaps at the $\Gamma$ point ($\Delta_\Gamma$) are summarized in the table in Fig. 4c. The narrowest energy gap of ~ 0.25 eV occurs at the BL edge and the ML-BL junction, whereas a slightly wider gap of 0.32 eV is seen at the ML edge. Larger gaps are obtained in the BL bulk (~ 1.15 eV) and ML bulk (~ 1.24 eV) regions. Since the DFT method usually underestimates the energy gap values, the theoretical calculation is in good qualitative agreement with our experimental observation.

Finally, the high quality of the MIM data allows us to quantitatively compare the photo-response between the ML and BL regions when the sample is illuminated by the above-gap $\lambda =$ 446 nm laser. Fig. 5a shows selected MIM-Re images under several laser intensities (complete set of data in Fig. S2). The MIM-Re and MIM-Im signals in the two bulk areas are plotted in Fig. 5b. Using the FEA modeling (Supporting Information Fig. S4), we can again obtain the Drude-like linear σ – P relation in Fig. 5c, here the sheet conductance $\sigma_{2D} = \sigma_{3D} \cdot t$. Compared with the ML region, the 20-times enhancement of photoconductivity in the BL region can be understood as follows. First, the IPCE of the BL is roughly twice as high as that of the ML, i.e., $\eta_{BL} = 2\eta_{ML} = 0.2\%$, due to the difference in thicknesses[36,37]. To obtain the information on carrier mobility, we fabricate ionic-liquid-gated devices on several monolayer and bilayer WSe$_2$ flakes for transport measurements. As shown in Fig. 5d, the measured hole mobility of the BL and ML flakes is about 33 and 9 $cm^2 \cdot V^{-1}s^{-1}$, respectively, consistent with previous studies[39]. Assuming that the majority photo-generated mobile carriers are p-type in WSe$_2$, i.e., $\tau_p \gg \tau_n$, we can estimate the lifetime in the bilayer and monolayer regions to be 30 ns and 10 ns, respectively. The longer carrier lifetime in the BL region than that in the ML region is a result of the indirect-to-direct gap transition[6], which is driven by the presence (absence) of interlayer coupling in bilayer (monolayer) WSe$_2$.

In summary, we report light-stimulated microwave imaging on CVD-grown monolayer-bilayer WSe$_2$ heterostructures. No electrodes are required and the measurement is completely noninvasive. The intrinsic photoconductivity of the bulk and edge of both layers, as well as the monolayer-bilayer heterojunction, is quantified under the illumination of excitation lasers with wavelengths ranging from the infrared to visible regime. The onset of photo-response provides an estimate of the local energy gaps in the five different regions, which is consistent with the DFT calculation. Quantitative analysis of the MIM data confirms the linear relation between the photoconductivity and the laser intensity, as well as the importance of interlayer coupling in the carrier recombination process. Our results highlight the opportunity to engineer nanoscale electronic bands in the TMD heterostructures, which is critically important for their optoelectronic applications.

## Methods

**Material synthesis.** The monolayer-bilayer WSe$_2$ heterostructure was grown by the chemical vapor deposition (CVD) method. The WO$_3$ powder (Sigma-Aldrich, 99.9%) was placed in a quartz boat located in the center of the heating zone in the furnace. The Se powder (Sigma-Aldrich, 99.99%) was placed in a separate quartz boat at the upstream, which was maintained at 265 °C during the reaction. The sapphire substrates (double side polished) were put at the downstream, where the Se and $WO_3$ vapors were brought to the sapphire substrates by an $Ar/H_2$ flowing gas. The center heating zone was heated to 900 °C and kept for 15 min for the growth of $WSe_2$ crystals. Finally, the furnace was cooled down to room temperature after the heater was turned off.

**STEM sample preparation.** The sample was coated by the polymethylmethacrylate (PMMA) A8 resist to support the film during the transfer process. To separate the film from the sapphire substrate, the sample was soaked in an HF solution ($HF:H_2O$ = 1:3) for 15 min. After rinsing with deionized water for several times, the sample was dipped into the water by the tweezer to release the film from the substrate. With the film floating on the water surface, we used a QUANTIFOIL holey carbon films on 400 Mesh Nickel TEM grid (Agar Scientific) to scoop the film. The PMMA-A8 polymer was then removed by acetone steam. The TEM grid with samples was finally baked in vacuum ($< 1 \times 10^{-6}$ torr) at 300 °C for overnight to remove the PMMA residue.

**STEM characterization.** HAADF-STEM imaging was conducted at 80 kV using a ThermoFisher USA (former FEI Co) Titan Themis Z (40-300kV) TEM equipped with a double Cs (spherical aberration) corrector, a high brightness electron gun (x-FEG) and Fischione STEM detector. The STEM probe was tuned for a semi-convergence angle of ~ 30 mrad and a probe current of approximately 50 pA with an inner collection angle of 60 mrad to operate the STEM under the Z contrast imaging conditions, where the contrast was proportional to $Z^\gamma$ (Z is the atomic number and $1.3 < \gamma < 2$), allowing W and Se to be easily distinguished. To reduce distortions due to the sample drift and to improve the signal-to-noise ratio, multiple (20 to 30) frames were acquired, post-aligned and summed afterwards. The image acquisition time per frame was set for 1 sec.

**Device fabrication and transport measurement.** Field-effect transistors (FETs) were fabricated on $WSe_2$ monolayer and bilayer flakes grown under the same condition as described above. The devices were prepared by the standard E-beam lithography (EBL) process using a PMMA 950 A4 (320 nm) photoresist. Before the exposure, a thick sputtered Au film (6 nm in thickness) was coated on the photoresist to avoid the severe charging effect on the insulating substrates. The first EBL was to make the alignment marker on sapphire and the second one was to pattern the source-drain and gate electrodes, followed by e-beam evaporation of Pd (20 nm)/Au (70 nm) and standard lift-off process by acetone. The FET transfer characteristics were measured under vacuum by a Keithley 4200 semiconductor parameter analyzer. The ionic liquid was used to gate $WSe_2$ and the p-type semiconductor behavior was observed.

**First-principles DFT calculations**. DFT calculations were performed using the Vienna Ab initio simulation package (VASP) with Generalized Gradient Approximation PBE pseudopotentials. The lattice constant is 3.32 Å and van der Waals gap for the bilayer $WSe_2$ is 6.4 Å. The AB stacking order is used in the calculation of bilayer $WSe_2$ structure, consistent with the STEM data. To model the edge states, we put a slab of vacuum (2 nm in width) next to the $WSe_2$ nanoribbon (2 nm in width) and repeat this unit cell perpendicular to the ribbon. For monolayer $WSe_2$, there are 6 atoms at the edge of the nanoribbon, and thus 12 sub-bands are identified as edge states inside the bulk band gap. The same configuration is used for bilayer $WSe_2$, except that there are 12 atoms at the edge and 24 sub-bands are identified as edge states. To model the ML-BL $WSe_2$ lateral heterostructure, we put a 1.66-nm-wide monolayer nanoribbon onto a second monolayer of $WSe_2$ ribbon (3.32 nm in width) and again repeat this structure periodically. The number of edge states is 12 in this case.

## Supporting Information

Raman maps of the $WSe_2$ sample in the main text; complete sets of the MIM images various wavelengths and power density; MIM data on a different flake; quantitative study of the MIM results by finite-element analysis. The Supporting Information is available free of charge on the ACS Publications website at xxx.

## Acknowledgements

This research performed collaboratively between K.L., X. L, and A. H. M was primarily supported by the National Science Foundation through the Center for Dynamics and Control of Materials: an NSF MRSEC under Cooperative Agreement No. DMR-1720595. X.L. and K.L. were also supported by NSF EFMA-1542747. The instrumentation was supported by the U.S. Army Research Laboratory and the U.S. Army Research Office under grants W911NF-16-1-0276 and W911NF-17-1-0190. K.L., Z.C. and D.W. acknowledge the support from Welch Foundation Grant F-1814. X. L. acknowledges support from Welch Foundation Grant F-1662. A.H.M. and C.L. acknowledge support from Welch Foundation Grant TBF1473. L.L. acknowledges the support from King Abdullah University of Science and Technology.

## Author contributions

K.L., X.L., and L.L. conceived the project. Z.C. carried out the optical and MIM measurements. A.H. grew and prepared the samples for all measurements. C.L. conducted the DFT calculation. S.L. and A.H. performed the electron microscopy and analyzed the STEM data. P.L. performed the transport measurement. Z.C. and K.L. performed the data analysis and drafted the manuscript. All authors have given approval to the final version of the manuscript.

## Conflict of interest

K.L. holds a patent on the MIM technology, which is licensed to PrimeNano Inc. for commercial instrument. The terms of this arrangement have been reviewed and approved by the University of Texas at Austin in accordance with its policy on objectivity in research. The remaining authors declare no competing financial interests.

## References

1. Novoselov, K. S., Mishchenko, A., Carvalho, A., Castro Neto, A.H., *Science* **353**, 461 (2016), aac9439.
2. Geim, A.K., Grigorieva, I.V. *Nature* **499**, (2013), 419-425.
3. Kang, J., Tongay, S., Zhou, J., Li, J., Wu, J. *Appl. Phys. Lett*. **102**, 012111 (2013).
4. Wang, H., Yuan, H., Hong, S.S., Li, Y., Cui, Y. *Chem. Soc. Rev*. **44**, (2015), 2664-2680.
5. Mak, K. F., Lee, C., Hone, J., Shan, J., Heinz, T. F. *Phys. Rev. Lett*. **105**, 136805 (2010).
6. Zhao, W., Ghorannevis, Z., Chu, L., Toh, M., Kloc, C., Tan, P.-H., Eda, G. *ACS Nano* **7(1)**, (2013), 791-797.
7. Fang, H., Battaglia, C., Carraro, C., Nemsak, S., Ozdol, B., Kang, J. S., Bechtel, H. A., Desai, S. B., Kronast, F., Unal, A. A., Conti, G., Conlon, C., Palsson, G. K., Martin, M. C., Minor, A. M., Fadley, C. S., Yablonovitch, E., Maboudian, R., Javey, A. *Proc. Natl. Acad. Sci*. **111**, (2014), 6198-6202.

8. Liu, K., Zhang, L., Cao, T., Jin, C., Qiu, D., Zhou, Q., Zettl, A., Yang, P., Louie, S.G., Wang, F. *Nature Commun*. **5**, 4966 (2014).
9. Tong, Q., Yu, H., Zhu, Q., Wang, Y., Xu, X., Yao, W. *Nature Phys*. **13**, (2017), 356-362.
10. Bollinger, M., Lauritsen, J., Jacobsen, K., Norskov, J., Helveg, S., Besenbacher, F. *Phys. Rev. Lett*. **87**, 196803 (2001).
11. Zhang, C., Johnson, A., Hsu, C., Li, L., Shih, C. *Nano Lett*. **14**, (2014), 2443-2447.
12. Qian, X., Liu, J., Fu, L., Li, J. *Science* **346**, (2014), 1344-1347.
13. Fei, Z., Palomaki, T., Wu, S., Zhao, W., Cai, X., Sun, B., Nguyen, P., Finney, J., Xu, X., Cobden, D. H. *Nature Phys*. **13**, (2017), 677-682.
14. Wu, S., Fatemi, V., Gibson, Q. D., Watanabe, K., Taniguchi, T., Cava, R. J., Jarillo-Herrero, P. *Science* **359**, (2018), 76-79.
15. Wang, Q. H., Kalantar-Zadeh, K., Kis, A., Coleman, J. N., Strano, M. S. *Nature Nanotech*. **7**, (2012), 699-712.
16. Mak, K., Shan, J. *Nature Photon.* **10**, (2016), 216-226.
17. Xia, F., Wang, H., Xiao, D., Dubey, M., Ramasubramaniam, A. *Nature Photon.* **8**, (2014) 899-907.
18. Bube, R. H. *Photoelectronic Properties of Semiconductors* (Cambridge University Press, Cambridge, UK, 1992).
19. Kam, K. K., Parkinson, B. A. *J. Phys. Chem*. **86**, (1982), 463-467.
20. Wu, C.-C., Jariwala, D., Sangwan, V. K., Marks, T. J., Hersam, M. C., Lauhon, L. J. *J. Phys. Chem. Lett*., **4**, (2013), 2508-2513.
21. Freitag, M., Low, T., Xia, F., Avouris, P. *Nature Photon.* **7**, (2013), 53-59.
22. Furchi, M., Polyushkin, D., Pospischil, A., Mueller, T. *Nano Lett*. **14**, (2014), 6165-6170.
23. Cheng, R., Li, D., Zhou, H., Wang, C., Yin, A., Jiang, S., Liu, Y., Chen, Y., Huang, Y., Duan, X. *Nano Lett*. **14**, (2014), 5590-5597.
24. Coffey, D. C., Reid, O. G., Rodovsky, D. B., Bartholomew, G. P., Ginger, D. S. *Nano Lett.* **7**, (2007) 738-744.
25. Lai, K., Kundhikanjana, W., Kelly, M. A. & Shen, Z. X. *Appl. Nanosci*. **1**, (2011), 13-18.
26. Tsai, Y., Chu, Z., Han, Y., Chuu, C., Wu, D., Johnson, A., Cheng, F., Chou, M., Muller, D., Li, X., Lai, K., Shih, C. *Adv. Mater*. **29**, 1703680 (2017).

27. Chu, Z., Yang, M., Schulz, P., Wu, D., Ma, X., Seifert, E., Sun, L., Li, X., Zhu, K., Lai, K. *Nature Commun*. **8**, 2230 (2017).
28. Tonndorf, P., Schmidt, R., Böttger, P., Zhang, X., Börner, J., Liebig, A., Albrecht, M., Kloc, C., Gordan, O., Zahn, D., Vasconcellos, S., Bratschitsch, R. *Opt. Express* **21**, (2013), 4908-4916.
29. Hsu, W., Lu, L., Wang, D., Huang, J., Li, M., Chang, T., Chou, Y., Juang, Z., Jeng, H., Li, L., Chang, W., *Nature Commun*.**8**, 929 (2017)
30. Chiu, K., Huang, K., Chen, C., Lai, Y., Zhang, X., Lin, E., Chuang, M., Wu, J., Lee, Y., *Adv. Mater*. 2018, 30, 1704796.
31. Krivanek, O., Chisholm, M., Nicolosi, V., Pennycook, T., Corbin, G., Dellby, N., Murfitt, M., Own, C., Szilagyi, Z., Oxley, M., Pantelides, S., Pennycook, S. *Nature* **464**, (2010), 571-574.
32. Yang, Y., Lai, K., Tang, Q., Kundhikanjana, W., Kelly, M. A., Zhang, K., Shen, Z.-X., Li, X. *J. Micromech. Microeng*. **22**, 115040 (2012).
33. Lai, K., Kundhikanjana, W., Kelly, M. & Shen, Z. X. *Rev. Sci. Instrum*. **79**, 063703 (2008).
34. Wu, D., Li, X., Luan, L., Wu, X., Li, W., Yogeesh, M., Ghosh, R., Chu, Z., Akinwande, D., Niu, Q., Lai, K. *Proc. Natl. Acad. Sci*. **113**, (2016), 8583-8588.
35. Zhang, C., Chen, Y., Huang, J., Wu, X., Li, L., Yao, W., Tersoff, J., Shih, C. *Nature Commun*. **7**, 10349 (2016).
36. Pospischil, A., Furchi, M., Mueller, T. *Nature Nanotech*. **9**, (2014), 257-261.
37. Gong, Y., Lei, S., Ye, G., Li, B., He, Y., Keyshar, K., Zhang, X., Wang, Q., Lou, J., Liu, Z., Vajtai, R., Zhou, W., Ajayan, P. M. *Nano Lett*. **15**, (2015), 6135-6141.
38. Liu, G.-B., Shan, W.-Y., Yao, Y., Yao, W., Xiao, D. *Phys. Rev. B* **88**, 085433 (2013).
39. Yu, C., Fan, M., Yu, K., Hu, V., Su, P., Chuang, C. *IEEE Trans. Electron Devices*. **63**, (2016), 625-630.

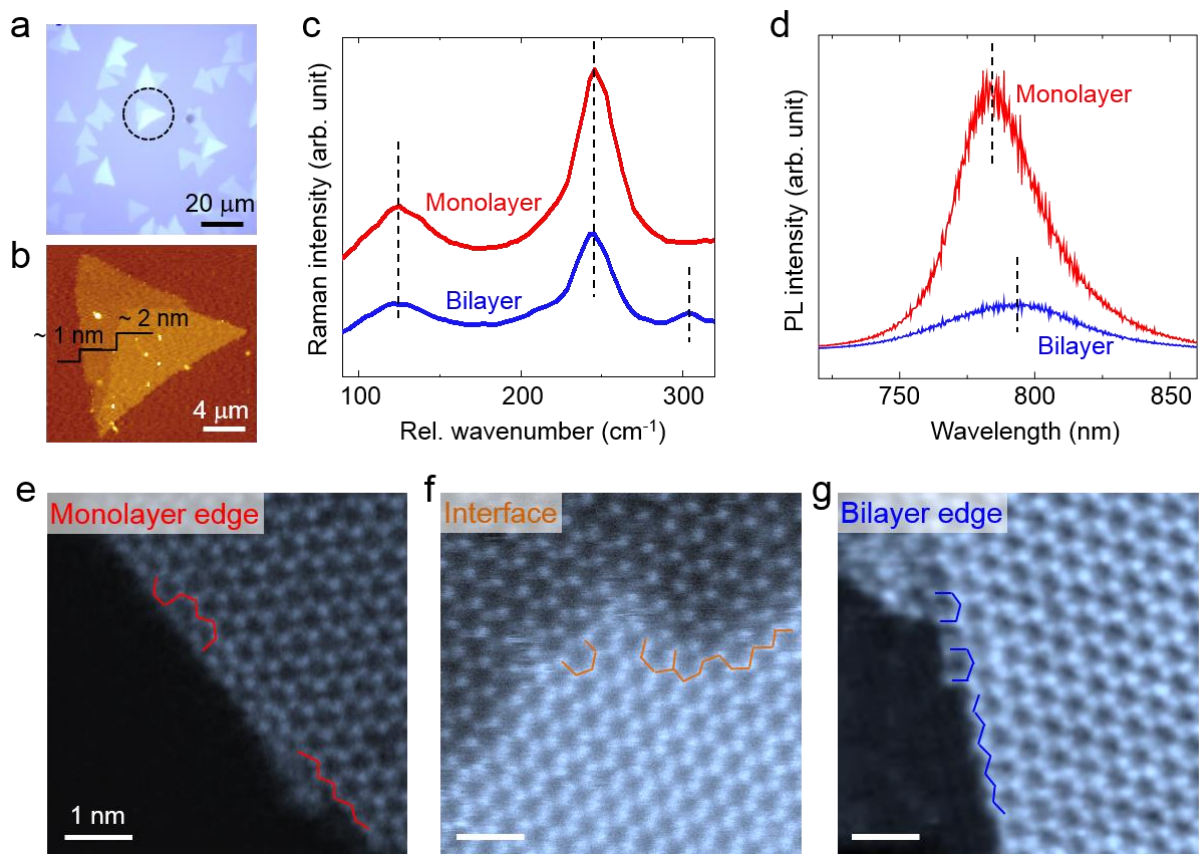

**Figure 1 | Characterization of the WSe$_2$ sample.** (**a**) Optical image of the CVD-grown WSe$_2$ sample, showing several flakes with the desired monolayer-bilayer heterostructure. (**b**) AFM image of the flake inside the dashed circle in (a). (**c**) Raman spectra in the monolayer and bilayer regions. (**d**) Photoluminescence spectra in the monolayer and bilayer regions. The dashed lines in (c) and (d) indicate the Raman and PL peaks. (**e** – **g**) HAADF-STEM images near the monolayer edge, monolayer-bilayer interface, and bilayer edge, respectively. Some zigzag and armchair sections are highlighted at the disordered edge and hetero-interface. The scale bars in (e – g) are 1 nm.

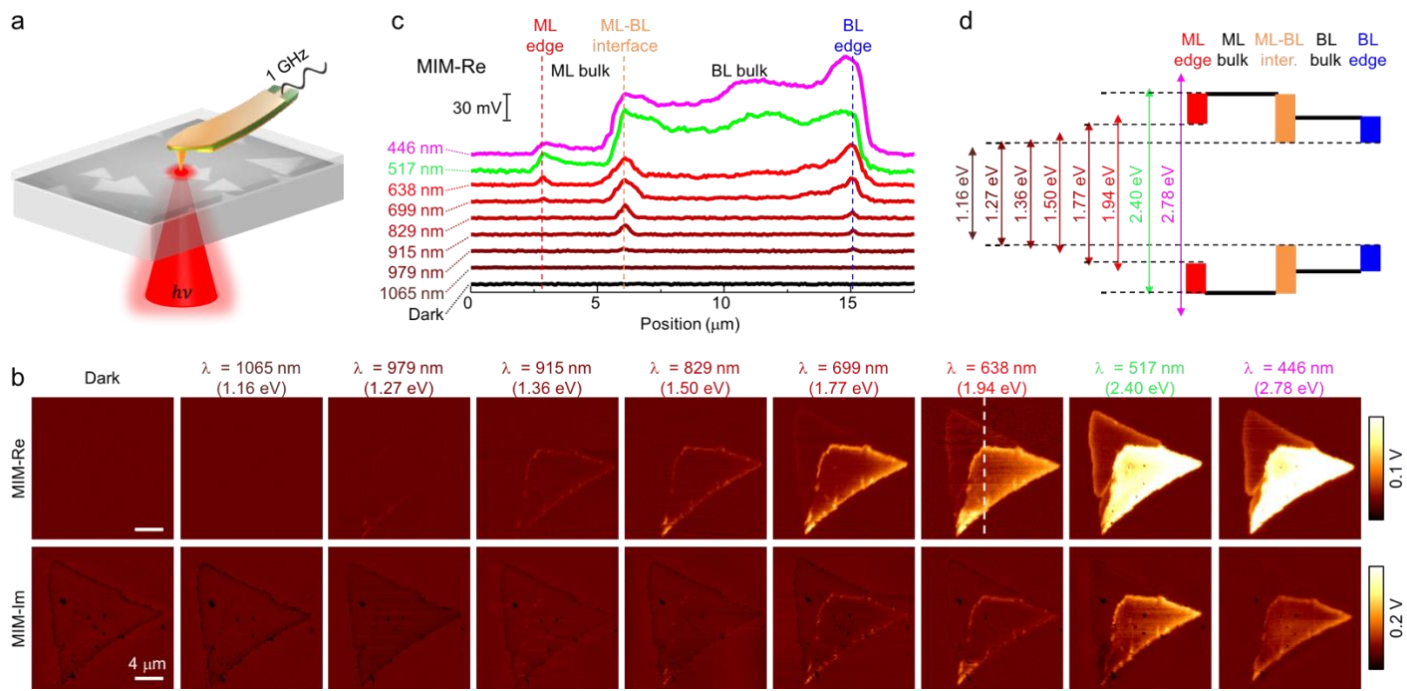

**Figure 2 | Wavelength-dependent photoconductivity imaging on monolayer-bilayer WSe$_2$ heterostructure**. (**a**) Schematic diagram of the light-stimulated MIM setup. (**b**) MIM images of the flake in Fig. 1b when illuminated by diode lasers with different energies. The scale bars are 4 μm. (**c**) Line profiles across the white dashed line in (b). Five regions in the sample (ML edge, ML bulk, ML-BL interface, BL bulk, and BL edge) are labeled in the plot. Micrometer-scale fluctuations in the ML and BL bulk regions are likely due to spatial variations of defects and strains. (**d**) Configuration of local energy gaps in the sample inferred from the MIM results.

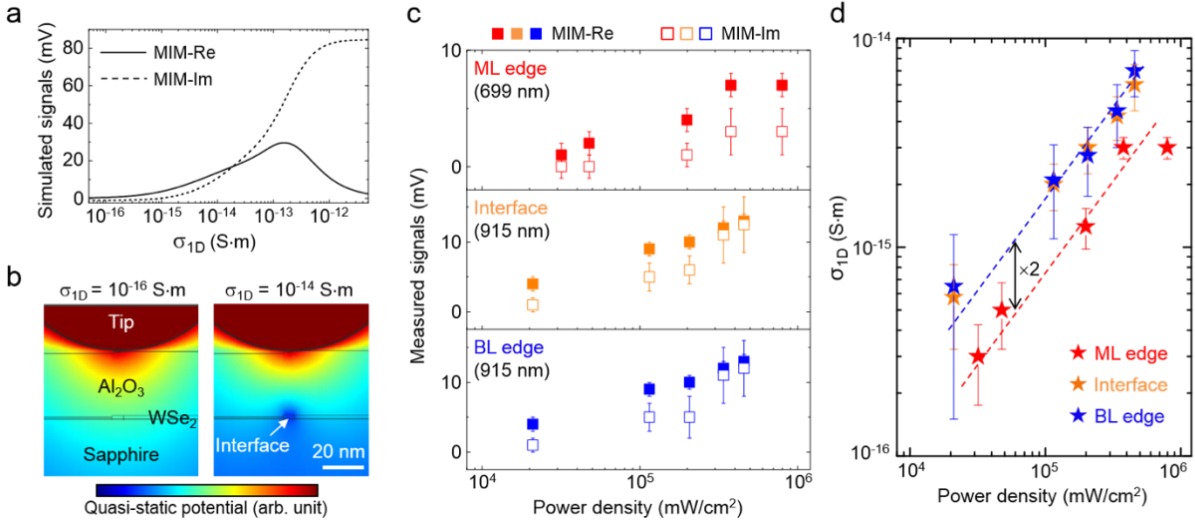

**Figure 3 | Quantitative analysis of photoconductivity at the hetero-interface and edges.** (**a**) Simulated MIM signals at the monolayer-bilayer interface as a function of the 1D conductance. (**b**) Quasi-static potential distribution around the tip apex at $\sigma_{1D} = 10^{-16}$ S·m (left) and $10^{-14}$ S·m (right), showing the screening effect due to mobile carriers. (**c**) Measured MIM signals as a function of the incident laser power for the monolayer edge (top, $\lambda = 699$ nm), interface (middle, $\lambda = 915$ nm) and bilayer edge (bottom, $\lambda = 915$ nm). The error bars are estimated by the standard deviation in the MIM images. (**d**) Photoconductivity of edges and interface as a function of power density. The dashed lines are linear fits to the data.

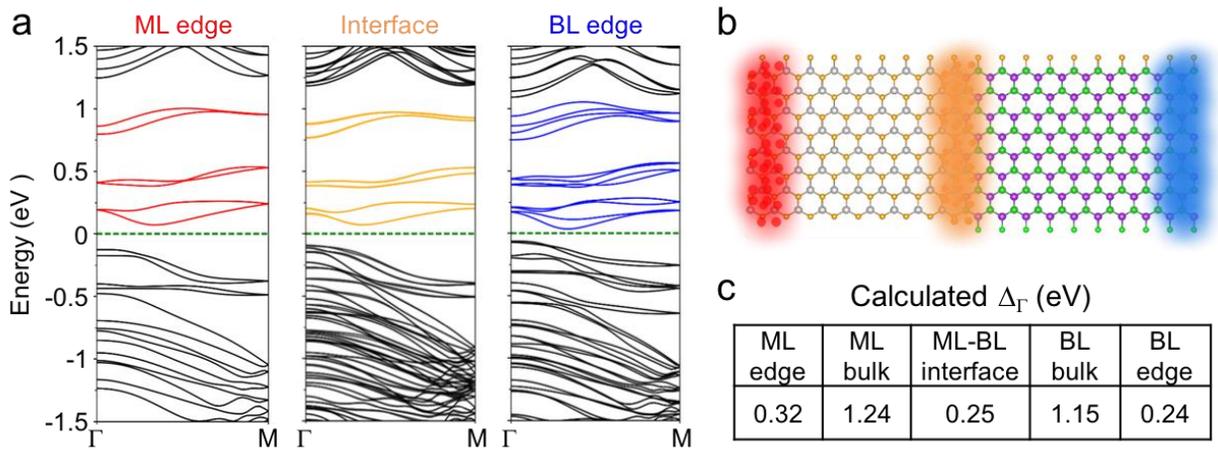

**Figure 4 | DFT calculations of local band structures.** (**a**) Calculated armchair configuration ribbon and interface electronic structure. The Fermi levels are denoted by green dashed lines. The electronic bands inside the bulk gaps are highlighted in red (monolayer edge), orange (monolayer-bilayer interface), and blue (bilayer edge) curves. These quasi-one-dimensional bands were calculated from left to right for 12-tungsten-atom-wide armchair ribbons, 20-tungsten-atom-wide supercells (with 20-tungsten-atom-wide bilayer and 10-tungsten-atom-wide monolayer) containing two monolayer/bilayer interfaces, and 12-tungsten-atom-wide bilayer ribbons. (**b**) Schematic diagram of edge and interfacial states in the real space. The color bands highlight the monolayer edge (red), interface (orange), and bilayer edge (blue). (**c**) Calculated band gaps at the $\Gamma$ point.

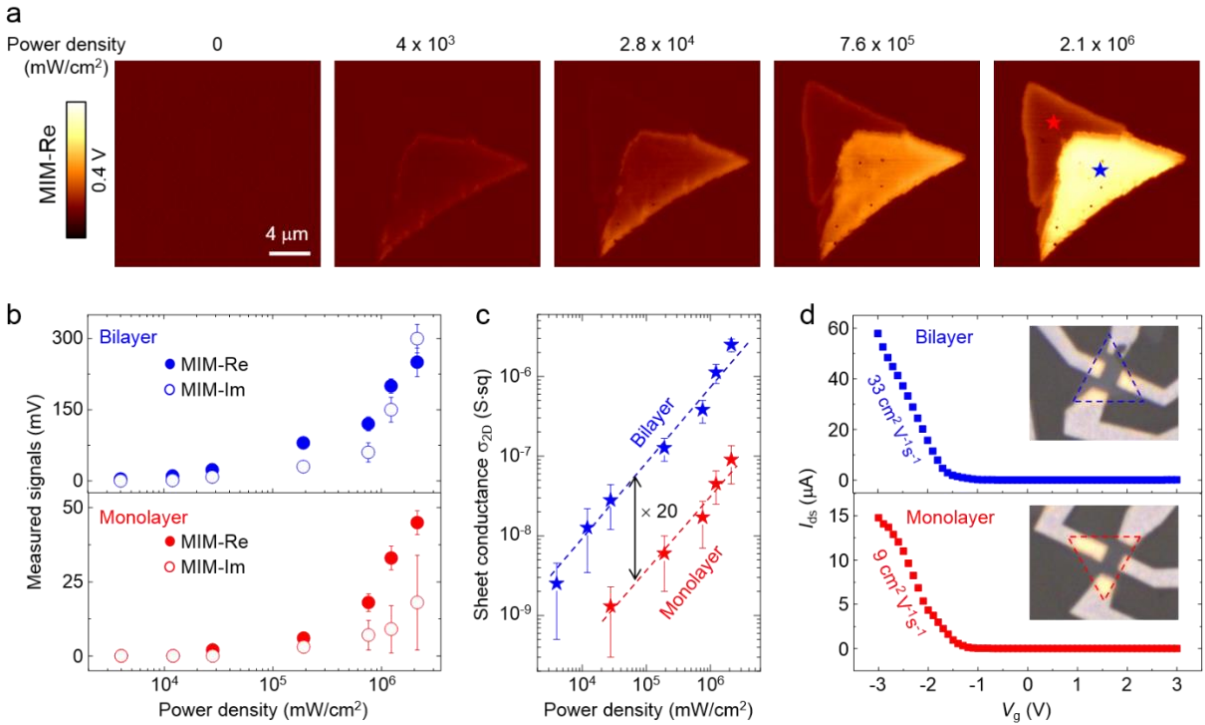

**Figure 5 | Power-dependent photoconductivity in monolayer and bilayer regions.** (**a**) MIM-Re images acquired under various laser ($\lambda$ = 446 nm, $E$ = 2.78 eV) intensities. (**b**) Average MIM signals of the bilayer (top) and monolayer (bottom) regions as a function of the laser intensity. (**c**) Power dependence of the sheet conductance of the monolayer (red) and bilayer (blue) bulk regions. Dashed lines are linear fits to the data. (**d**) Transfer curves of the bilayer (top) and monolayer (bottom) WSe$_2$ FET devices. The mobility of bilayer and monolayer flakes is measured to be 9 and 33 cm$^2$V$^{-1}$s$^{-1}$, respectively. The insets are the optical images of the devices.